# Symmetry-driven giant magneto-optical Kerr effects in altermagnet hematite

Jiaxin Luo[1,2,*], Xiaodong Zhou[3,*], Jinxuan Liang[1,2,*], Ledong Wang[1,2], Qiuyun Zhou[1,2], Yong Jiang[4], Wenhong Wang[4], Yugui Yao[5,6], Luyi Yang[1,2,‡], Wanjun Jiang[1,2,‡]

[1] *State Key Laboratory of Low-Dimensional Quantum Physics and Department of Physics, Tsinghua University, Beijing 100084, China*

[2] *Frontier Science Center for Quantum Information, Tsinghua University, Beijing 100084, China*

[3] *School of Physical Science and Technology, Tiangong University, Tianjin 300387, China*

[4] *School of Electronic and Information Engineering, Tiangong University, Tianjin 300387, China*

[5] *Centre for Quantum Physics, Key Laboratory of Advanced Optoelectronic, Quantum Architecture and Measurement (MOE), School of Physics, Beijing Institute of Technology, Beijing 100084, China*

[6] *Beijing Key Lab of Nanophotonics and Ultrafine Optoelectronic Systems, Beijing Institute of Technology, Beijing 100084, China*

[*] These authors contributed equally.

[†] Correspondence should be addressed to: luyi-yang@mail.tsinghua.edu.cn

[‡] Correspondence should be addressed to: jiang_lab@tsinghua.edu.cn

**Abstract**

Altermagnets have attracted tremendous interest for revealing intriguing physics and promising spintronics applications. In contrast to conventional antiferromagnets, altermagnets break both $PT$ and $T\tau$ symmetries, and simultaneously exhibit spin-split band structures with a vanishing net magnetization. To quantify insulating altermagnets without conduction electron, we propose to use magneto-optical Kerr effect (MOKE) to identify the altermagnetic fingerprints. In particular, we demonstrate not only the giant MOKE responses, but also their connection with the orientations of Néel vectors at room temperature in altermagnet hematite ($\alpha\text{-Fe}_2\text{O}_3$). Specifically, under the Néel vector along the $[\bar{1}100]$ axis, we find a giant polar Kerr rotation angle 93.4 mdeg in the $(11\bar{2}0)$ plane, which is allowed by the magnetic space group $C2'/c'$. Under the Néel vector along the $[11\bar{2}0]$ axis, we find a longitudinal Kerr angle 9.6 mdeg in the $(0001)$ plane, which is allowed by the magnetic space group $C2/c$. Further, we show that such pronounced MOKE effects directly enable an optical imaging of altermagnetic domains, together with their reversible domain wall (DW) motion. Our studies not only suggest MOKE can be used to identify altermagnet candidates, but also signify the feasibility of exploring altermagnetic optical and DW spintronics, which could largely expand the current research paradigm of altermagnetism.

**Introduction**

Symmetry breaking plays an important role in many topics in condensed matter physics [1], and recently allows a new class of magnet – altermagnet to be identified [2,3]. The concept of altermagnet is attributed to the broken $PT$ and $T\tau$ symmetries, together with a compensated sublattice magnetization [2,3], where $P$ is the inversion, $T$ is the time-reversal, and $\tau$ is the half-unit cell translation operation. In particular, altermagnets feature spin-split band structures with a zero magnetization [4–9], which is accompanied with unconventional spin transport properties, including the anomalous Hall effect (AHE) [10–21], anomalous Nernst effect [22–24], spin splitting torque [25–27], and other related spintronic effects [28–32]. Such methods can be conveniently used to study metallic altermagnets, which are however inapplicable to insulating materials due to the absence of conduction electron.

Note that magneto-optical Kerr effects (MOKE) could sense the time-reversal symmetry breaking in quantum materials, including unconventional superconductors [33,34] and magnetic topological materials [35,36], which is enabled by the difference in the refractive index between left and right circularly polarized lights [37]. However, the implementation of MOKE for identifying the altermagnetic signatures requires thorough considerations of the magnetic crystal symmetry [10,11,38,39]. Specifically, the ac Hall pseudovector $\boldsymbol{\sigma}_H = \left(\sigma_{yz}(\omega), \sigma_{zx}(\omega), \sigma_{xy}(\omega)\right)$ governs the physics of MOKE, where $\sigma_{ij}$ are the off-diagonal optical conductivity components and $i,j,k \in [x,y,z]$. The $\boldsymbol{\sigma}_H$ transforms like a magnetic dipole moment under symmetry operations, and is odd under $PT$ and $T\tau$ operations [11]. In antiferromagnets, the naturally preserved $PT$ and $T\tau$ symmetries thus lead to $\boldsymbol{\sigma}_H = 0$ and the absence of MOKE [10,11,38]. By contrast, the broken $PT$ and $T\tau$ symmetries in altermagnets could make $\boldsymbol{\sigma}_H \neq 0$ and the presence of MOKE responses, as illustrated in Fig. 1. The specific nonzero components of $\boldsymbol{\sigma}_H$ can be identified from their invariance under operations of the specific magnetic space group (MSG) [10,11,38,40]. More importantly, the MSG of altermagnets can be changed by switching the orientation of the Néel vector [10,11,41]. Thus, Néel-vector-orientation-dependent MOKE responses can be used to validate the magnetic crystal symmetry in altermagnets, which is of particular importance for studying insulating altermagnets where electron transport is inapplicable.

Recently, hematite ($\alpha$-$Fe_2O_3$), one of the most well-known antiferromagnets [42–47], has been theoretically relabeled as a g-wave altermagnet candidate with a spin Laue group $^1\bar{3}^2$m [2,3,11,21,41,48]. Such an insulating candidate provides a suitable platform to examine the symmetry-driven MOKE responses, which is also important to reinvigorate the foundation of altermagnetism. Following these motivations, we experimentally study the MOKE responses in the $(11\bar{2}0)$-plane and the $(0001)$-plane terminated hematite single crystals at room temperature. Through verifying the connection between the orientations of Néel vectors and the MOKE responses, we identify the critical role of the symmetry in MSGs of altermagnets. We further show the observation of altermangetic domains and their field-driven domain wall (DW) motion by MOKE imaging microscopy. The incorporation of MOKE could accelerate the research on altermagnetic materials, physics and device concepts.

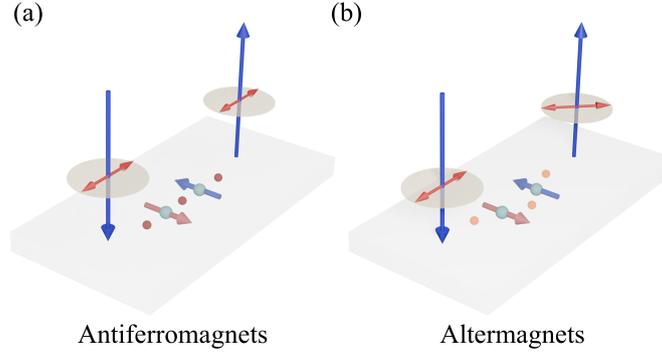

**FIG. 1. MOKE in antiferromagnets and altermagnets.** (a) In conventional collinear antiferromagnets, there are no MOKE responses. (b) In altermagnets, the broken $PT$ and $T\tau$ symmetries make MOKE possible.

## Results

**Magnetic crystal symmetry-driven MOKE in hematite.**

Hematite exhibits a hexagonal crystal structure (space group: $R\bar{3}c$), as shown in Fig. 2(a). It undergoes the Morin transition ($T_M$) at ~263 K [42,49]. Below $T_M$, hematite is in the easy-axis phase with spins aligning along the [0001] axis. Above $T_M$, the spins reorientate into the (0001) plane, which transforms hematite into the easy-plane phase. As shown in Fig. 2(a), spins are antiferromagnetically coupled along the [0001] axis. The rhombohedral primitive unit cell in Fig. 2(b) clearly shows the broken $T\tau$ and $PT$ symmetries. More importantly, the oppositely aligned sublattices can be transposed by three glide mirrors $[C_2||M_{[1\bar{1}00],[\bar{1}010],[01\bar{1}0]}|\tau]$ [41], in which the operation on the left of $||$ is in spin space, whereas the right part denotes operations in the real space. As a result, the spin degeneracy occurs within the $M_{[1\bar{1}00],[\bar{1}010],[01\bar{1}0]}$ glide mirror planes and the electronic bands show the alternating spin splitting between these planes in the $\boldsymbol{k}$-space (Supplementary Note 1). Thus, hematite could be considered as an altermagnet [2,3,11,41,48].

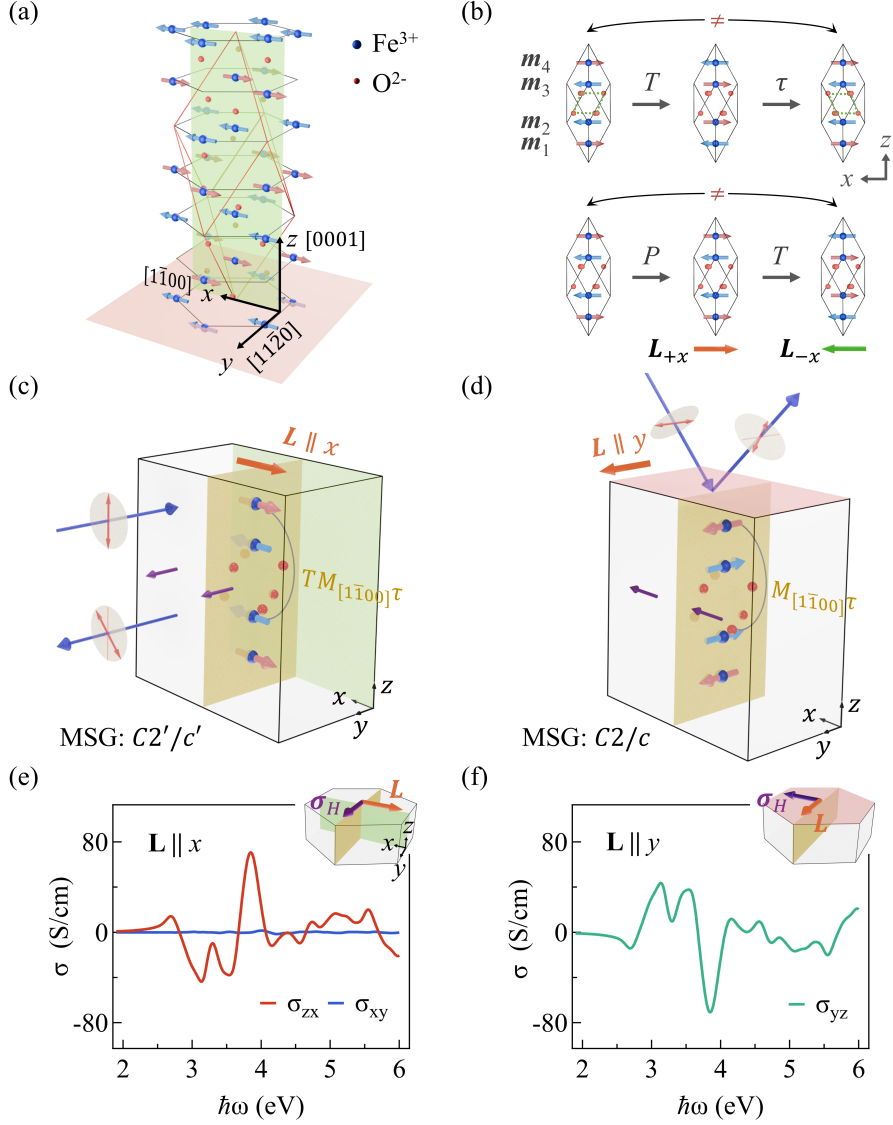

**FIG. 2. Magnetic crystal symmetry-driven MOKE in hematite.** (a) The hexagonal unit cell contains a rhombohedral primitive cell, which is marked in red. The light green plane ($xz$ plane) marks the surface of the $(11\bar{2}0)$-plane hematite, and the light pink plane ($xy$ plane) marks the surface of the (0001)-plane hematite. The Cartesian coordinates $x/y/z$ correspond to crystallographic directions $[1\bar{1}00]/[11\bar{2}0]/[0001]$, which remains throughout our studies. (b) The illustration shows the broken $T\tau$ and $PT$ symmetries, where $\tau = (1/2, 1/2, 1/2)$ is a half-unit cell lattice translation in the rhombohedral primitive cell. Here, $\boldsymbol{L}_{+x}$, $\boldsymbol{L}_{-x}$ indicate the orientations of two opposite Néel vectors, which point to the $[\bar{1}100]$ and the $[1\bar{1}00]$ axes, respectively. (c) $\boldsymbol{L} \parallel x$,

the yellow plane (($1\bar{1}00$) plane) preserves the sublattices transposing symmetry of $TM_{[1\bar{1}00]}\tau$. Purple arrows are the components of $\boldsymbol{\sigma}_H$. (d) $\boldsymbol{L} \parallel y$, the ($1\bar{1}00$) plane preserves symmetry of $M_{[1\bar{1}00]}\tau$. (e-f) The theoretical calculation of MSG-allowed components of $\boldsymbol{\sigma}_H = (0, \sigma_{zx}, \sigma_{xy})$ for $\boldsymbol{L} \parallel x$ and $\boldsymbol{\sigma}_H = (\sigma_{yz}, 0, 0)$ for $\boldsymbol{L} \parallel y$, respectively. The purple arrows in the insets are the MSG-allowed $\sigma_{ij}$ with respect to Néel vectors $\boldsymbol{L} \parallel x$ (only the dominant $\sigma_{zx}$ component is shown) and $\boldsymbol{L} \parallel y$. The yellow plane in the insets is the ($1\bar{1}00$) plane.

The complex Kerr rotation angle ($\tilde{\Theta}_k(\omega)$) can be calculated from [40]:

$$\tilde{\Theta}_k(\omega) = \theta_k + i\eta_k = -\frac{v_{ijk}\sigma_{ij}(\omega)}{\sigma_0(\omega)\sqrt{1+i(4\pi/\omega)\sigma_0(\omega)}} \quad \text{Eq. (1),}$$

where $\theta_k$ and $\eta_k$ denote the Kerr rotation angle and ellipticity, $v_{ijk}$ is the Levi-Civita symbol, and $\sigma_0 = (\sigma_{ii} + \sigma_{jj})/2$ ($i \neq j$). The Kerr rotation angle $\theta_k$ that is governed by $\sigma_{ij}$ can be detected with different MOKE configurations. In the primitive unit cell, the Néel vector is defined as $\boldsymbol{L} = \boldsymbol{m}_1 - \boldsymbol{m}_2 - \boldsymbol{m}_3 + \boldsymbol{m}_4$, as shown in Fig. 2(b) [50]. In particular, the rotation of the Néel vector $\boldsymbol{L}$ within the (0001) plane (easy plane) leads to a periodic change of the MSG, thereby accordingly modifying the MOKE responses. The MSG-allowed $\sigma_{ij}$ components for different orientations of Néel vectors are summarized in Table S1.

Since spins are axial vectors, the switching of Néel vectors with respect to mirror planes can change the magnetic crystal symmetry and the corresponding MSG in altermagnets [10,41,51]. Two orthogonal orientations of Néel vectors ($\boldsymbol{L} \parallel x$ and $\boldsymbol{L} \parallel y$) are chosen to probe the symmetry-allowed MOKE responses, as shown in Figs. 2(c-f). For the case of $\boldsymbol{L} \parallel x$ in Fig. 2(c), spins are perpendicular to the ($1\bar{1}00$) plane, so that the two opposite sublattices are transposed under the $TM_{[1\bar{1}00]}\tau$ operations (MSG $C2'/c'$). As a result, specific components of pseudovector $\boldsymbol{\sigma}_H$ ($\sigma_{zx}$ and $\sigma_{xy}$), which are parallel with the ($1\bar{1}00$) plane, are permitted (Supplementary Note 1). The dominant $\sigma_{zx}$ component from the calculation in Fig. 2(e) can be conveniently examined via polar MOKE in the $(11\bar{2}0)$-plane hematite. For the case of $L \parallel y$ in Fig. 2(d), the two opposite sublattices are transposed under the $M_{[1\bar{1}00]}\tau$ operation (MSG $C2/c$), due to spins being parallel with the $(1\bar{1}00)$ plane. The survival component of $\boldsymbol{\sigma}_H$ ($\sigma_{yz}$) is perpendicular to the $(1\bar{1}00)$ plane. By contrast, the $\sigma_{xy}$ and $\sigma_{zx}$ are forbidden. The presence of $\sigma_{yz}$ for $L \parallel y$ can be measured via longitudinal MOKE, whereas the absence of $\sigma_{xy}$ can be revealed via polar MOKE on the same $(0001)$-plane hematite. The illustration of MOKE configurations and termination planes of hematite, which measure different components of $\sigma_{ij}$, can be found in Supplementary Note 1.

**MOKE in hematite at room temperature.**

In the easy-plane phase at room temperature, the canted magnetization within the $(0001)$ plane is measured as $\boldsymbol{m}_{DM} \sim 2.1$ emu/cc, which corresponds to a spin canting angle $\varphi \sim 0.065°$ arising from the Dzyaloshinskii–Moriya interaction (DMI) in hematite [41,52,53]. Due to the parallel configuration between $\boldsymbol{m}_{DM}$ and magnetic field $\boldsymbol{H}$ within the $(0001)$ plane ($xy$ plane), the switching of Néel vector $\boldsymbol{L}$ by external field $\boldsymbol{H}$ can be understood from the form of DM energy:

$$\Delta E_{DM} \propto -\boldsymbol{D} \cdot (\boldsymbol{L} \times \boldsymbol{H}) \qquad \text{Eq. 2,}$$

where the DM vector $\boldsymbol{D}$ is along the $[000\bar{1}]$ direction [41,54,55]. As shown in Fig. 3(a), DMI energy favors $\boldsymbol{L} \times \boldsymbol{H}$ being parallel with $\boldsymbol{D}$ and suggests the feasibility of switching the orientation of the Néel vector $\boldsymbol{L}$ by $\boldsymbol{H}$. Consequently, opposite directions of $\boldsymbol{H}_\pm$ lead to the opposite directions of $\boldsymbol{L}_\pm$ (Supplementary Note 2).

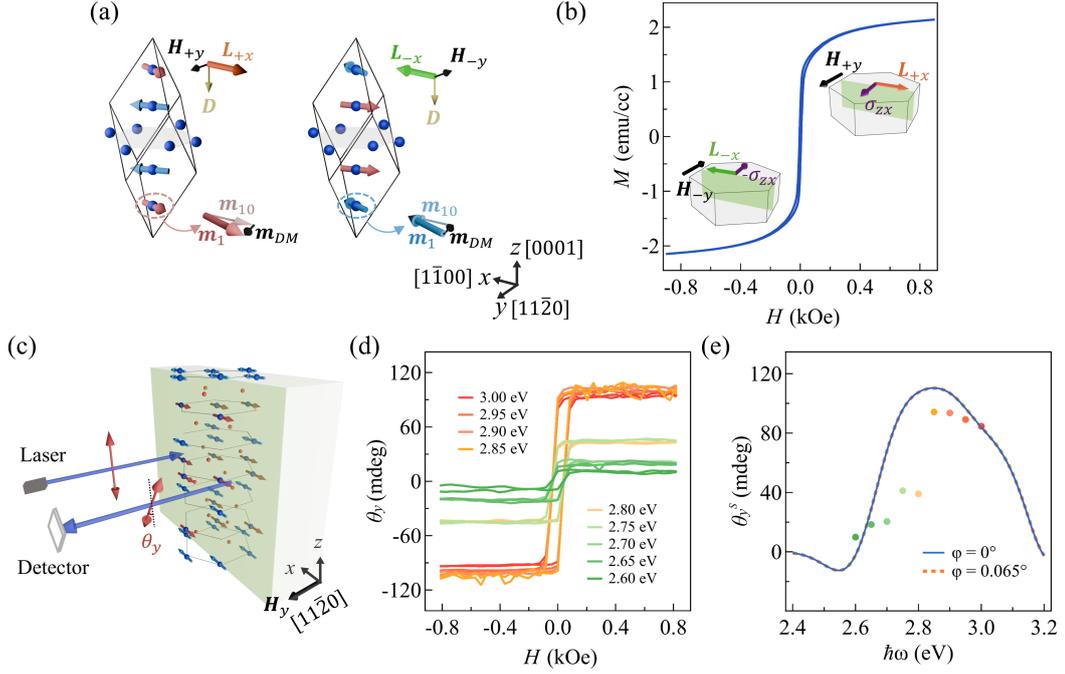

**FIG. 3. Polar MOKE in the $(11\bar{2}0)$-plane hematite with $L \parallel x$ at room temperature.** (a) Insets illustrate the switching of Néel vectors ($L_{\pm x}$) by the external field ($H_{\pm y}$) in the presence of the weak canted magnetization $m_{DM}$. Here, $m_1 = m_{10} + m_{DM}$ and $m_{10}$ is the projection of $m_1$ along the $x$ axis. (b) The canted magnetization $M$ is measured with $H \parallel y$. $M$ contains both the intrinsic $m_{DM}$ and field-induced canted magnetization along the field direction in this work. (c) Schematic illustration of the polar MOKE configuration on the $(11\bar{2}0)$-plane hematite. (d) Field dependence of the polar Kerr rotation angle $\theta_y$ under different photon energies. (e) The colored dots represent the spontaneous (zero field) polar Kerr rotation angles $\theta_y^s$. The calculated polar MOKE rotation spectra are given with (w) or without (w/o) incorporating the canting angle $\varphi$. The calculated curves are scaled by a factor of 1/2 for a clear comparison with the experimental results.

As shown in Fig. 3(b), the canted magnetization $M$ of the $(11\bar{2}0)$-plane hematite exhibits a clear hysteresis loop, which originates from the switching of $m_{DM}$ by magnetic field along the $y$ axis ($H_y$). Due to the cross configuration between $m_{DM}$ and $L$, such a switching of $m_{DM}$ is accompanied by the switching of $L_{\pm x}$. Since $H_y$ is perpendicular to the sample surface, we could use the polar MOKE at normal incidence to measure the $\sigma_{zx}$ on the $(11\bar{2}0)$-plane hematite, as shown in Fig. 3(c). In Fig. 3(d), the polar Kerr rotation angle ($\theta_y$) that measures $\sigma_{zx}$ exhibits clear hysteresis loops as a function of $H_y$ with the photon energy ranging from 2.60 eV to 3.00 eV. The spontaneous polar Kerr rotation angles $\theta_y^s$ at each photon energy are extracted at zero field with $\theta_y^s = (\theta_{y,+0\,\text{Oe}} + \theta_{y,-0\,\text{Oe}})/2$ and indicated by colored dots in Fig. 3(e). In particular, we can identify a pronounced $\theta_y^s$ around 94.3 mdeg at 2.85 eV. Compared to the small coercive field 6 Oe of the *M-H* loop in Fig. 3(b), the Kerr loops in Fig. 3(d) exhibit larger coercive fields around 40 Oe, which is likely due to the pinning effect of local domains rather than the volume-averaged bulk signal, together with the surface sensitivity of MOKE.

Note that $\theta_y$ remains nearly saturated (> 0.12 kOe), as shown in Fig. 3(d). By contrast, the field-induced canted magnetization $M$ increases with the magnetic field ($H_y$) in Fig. 3(b). This phenomenon is more pronounced in a wider field range from 0.45 kOe to 11 kOe, as shown in Supplementary Note 3. There, one can observe that $\theta_y$ remains nearly constant (a fluctuation of 3.7%), in stark contrast to the 63.4% monotonic increase of the canted magnetization $M$ in the same field range. We can conclude that the polar MOKE in hematite is not dominated by the canted magnetization. This aspect can be confirmed by theoretically calculating $\theta_y$ with/without incorporating the contribution from the canted magnetization $M$, as shown in Fig. 3(e). Specifically, the spontaneous $\theta_y^s$ shows a negligible difference between the non-canted case (canting angle $\varphi = 0°$) and the canted case ($\varphi = 0.065°$). Such a comparison suggests that the polar MOKE response primarily originates from the switching between the opposite Néel vectors $L_{+x}$ and $L_{-x}$ but not the canted magnetization. Note that theoretical calculations likely overestimate the amplitude of $\theta_y^s$, possibly due to the disorder [56,57], band broadening or spin fluctuations [58,59], which are not considered in our calculations.

Considering the minimal canted magnetization $M$ of 2.1 emu/cc (at 0.8 kOe in Fig. 3(b)), it is interesting to note that the observed spontaneous Kerr rotation angle $\theta_y^s$ (93.4 mdeg) is comparable with typical ferromagnets (535 mdeg in Fe, 145 mdeg in Ni [60]) and noncollinear/noncoplanar antiferromagnets [61–65] (20 mdeg in Mn$_3$Sn [66]). Note that polar MOKE with $L \parallel x$ on the (0001)-plane hematite is also conducted, which reveals a negligible $\theta_z$, as shown in Supplementary Note 4. This is also consistent with the predicted minimal amplitude of $\sigma_{xy}$, which is shown as the blue curve in Fig. 2(e).

To examine the connection between the orientation of the Néel vector and the MOKE response, we also study the case of $L \parallel y$, with the $M$-$H$ loop being shown in Fig. 4(a). In this case, $H \parallel x$ could flip the canted magnetization $M$, which simultaneously switches the Néel vector from $L_{+y}$ along the $[11\bar{2}0]$ axis to $L_{-y}$ along the $[\bar{1}\bar{1}20]$ axis, as suggested by Eq. 2. As discussed in Figs. 2(d) and 2(f), $\sigma_{yz}$ is allowed by MSG ($C2/c$) with respect to $L \parallel y$, which can be measured by longitudinal MOKE on the (0001)-plane hematite under $H \parallel x$, as shown in the inset of Fig. 4(b). In particular, a hysteresis loop of $\theta_x$ corresponding to $\sigma_{yz}$ is obtained in Fig. 4(b), in which a spontaneous longitudinal Kerr rotation angle $\theta_x^s = 9.6$ mdeg can be found.

Based on the symmetry analysis of MSG ($C2/c$) with $L \parallel y$ in Figs. 2(d) and 2(f), the absence of $\theta_z$ corresponding to $\sigma_{xy}$ is examined by polar MOKE measurements on the same (0001)-plane hematite, as shown in Figs. 4(c-d). Here, the $M$-$H$ and $\theta_z$-$H$ loops are performed with a tilted magnetic field within the $(11\bar{2}0)$ plane ($xz$ plane). The reduced magnetization (around 0.58 emu/cc) is consistent with the projected field along the $x$ axis. Under this configuration, the projected field along the $x$ axis ($H_{\pm x}$) flips $m_{DM}$ and consequently the directions of Néel vectors from $L_{+y}$ towards $L_{-y}$. However, no detectable polar MOKE response of $\theta_z$ can be found. Combining the theoretical and experimental results for $L \parallel x$ and $L \parallel y$ cases, one can conclude that MOKE in hematite is jointly determined by the orientation of the Néel vector and the symmetry of MSG of altermagnets.

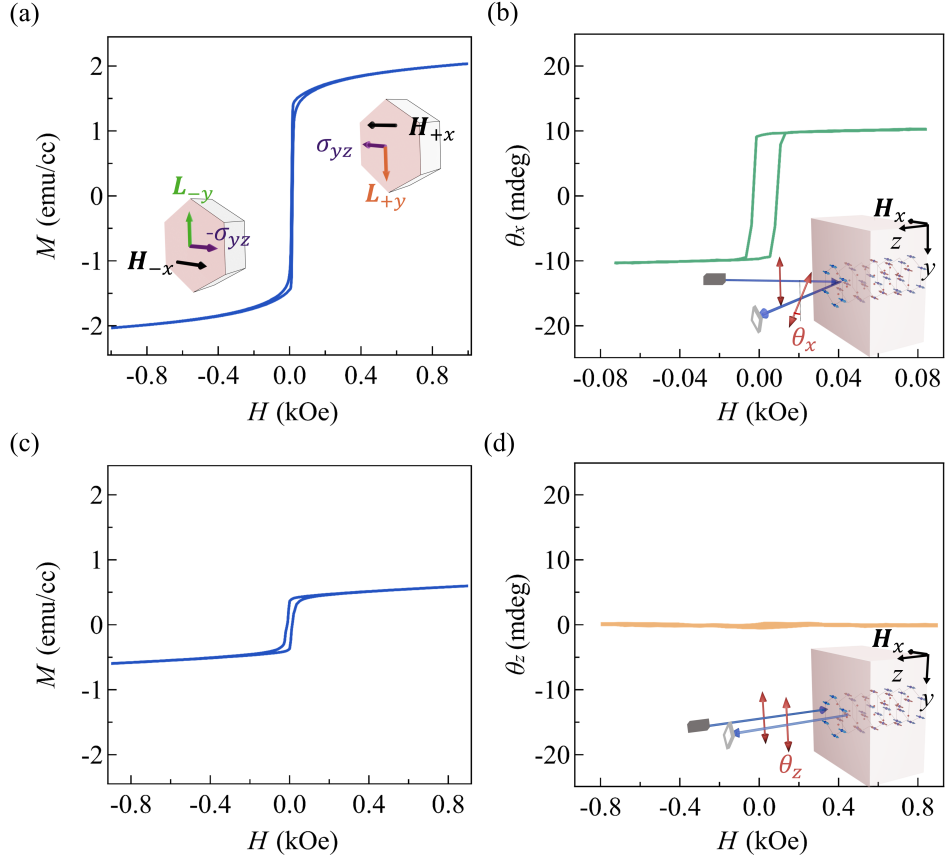

**FIG. 4. MOKE in the (0001)-plane hematite with $L \parallel y$.** (a) Field dependence of the canted magnetization $M$ with $H \parallel x$ in the (0001)-plane hematite (the sample surface is highlighted in light pink). (b) The field dependence of longitudinal Kerr rotation angle $\theta_x$ is obtained under 2.90 eV with $H \parallel x$. (c) Dependence of the canted magnetization $M$ on the magnetic field that is applied within the $(11\bar{2}0)$ plane ($xz$ plane) (approximately 16° away from the $z$ axis). (d) The absence of the polar Kerr rotation angle $\theta_z$ under 2.90 eV with $H$ in the $xz$ plane. Only the $H_x$ component is labeled in the inset for clarity. The inset is the schematic illustration of polar MOKE configuration on the (0001)-plane hematite.

**MOKE imaging of altermagnetic domains and their dynamics.**

Imaging studies of antiferromagnets [67–69] and altermagnets [21,70,71] mainly rely on the synchrotron-based and other specific techniques. Given the large polar MOKE responses in the $(11\bar{2}0)$ plane hematite, we try to optically image the altermagnetic domains and their dynamics by using MOKE imaging microscopy [37]. In Fig. 5(a), a clear polar MOKE contrast obtained in the $(11\bar{2}0)$ plane reflects the difference between two altermagnetic domains, which are separated by a DW. The MOKE contrast corresponds to the opposite orientations of Néel vectors $\boldsymbol{L_{+x}}$ and $\boldsymbol{L_{-x}}$. In the easy-plane phase, the field-driven altermagnetic DWs motion can be understood from the damping-like torque on the moments inside the DW [72]:

$$\boldsymbol{T}_{DW} \propto -\boldsymbol{m}_{DM}^{DW} \times \left(\boldsymbol{m}_{DM}^{DW} \times \boldsymbol{H}_y\right) \qquad \text{Eq. 3.}$$

Here, $\boldsymbol{m}_{DM}^{DW}$ is the magnetization in the center of the altermagnetic DW, which is along the $x$ axis (assuming the moments inside the DW are within the easy plane). With the increase of $\boldsymbol{H}_y$, $\boldsymbol{T}_{DW}$ rotates $m_{DM}^{DW}$ towards $\boldsymbol{H}_y$, which induces the motion of DW until the sample saturates with the energetically favored Néel vector direction. Thus, by sweeping $\boldsymbol{H}_y$, one could optically visualize the field-driven DW motion, as shown in Fig. 5(b). Through conducting the polar MOKE imaging experiment at different fields, the evolution of MOKE intensity as a function of $\boldsymbol{H}_y$ is obtained, and the threshold field (~12.2 Oe) between the reversible and irreversible motion of DW is also determined. These are shown in Supplementary Note 5.

To reveal the reversible and deterministic motion of altermagnetic DWs, we conduct the polar MOKE imaging experiment by applying a square wave of $\boldsymbol{H}_y$ with an amplitude of 1 Oe and a frequency of 0.1 Hz. Shown in Fig. 5(b) are the alternatively reversed time-dependent polar MOKE signals that are obtained from the same region (white circles in the insets) in the corresponding polar MOKE images. Such a deterministic contrast reversal not only suggests the reversible switching between $\boldsymbol{L_{+x}}$ and $\boldsymbol{L_{-x}}$, but also mimics the typical performances of DW memory devices.

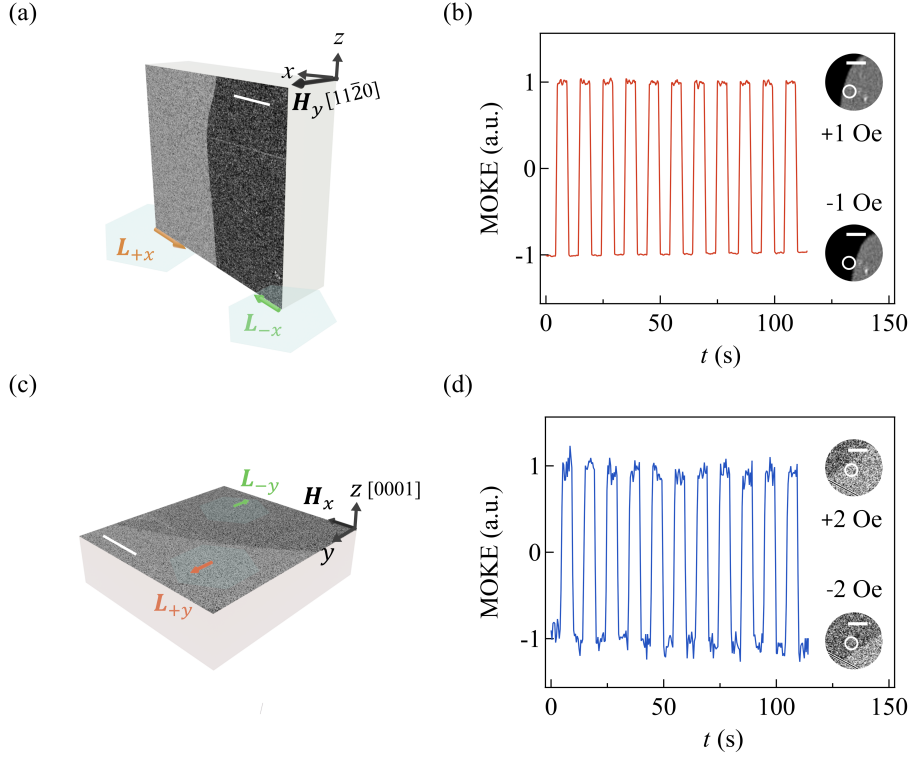

**FIG. 5. MOKE microscopy imaging of altermagnetic domains in hematite.** (a)/(c) The observation of two time-reversal altermagnetic domains by using polar/longitudinal MOKE imaging microscopy in the $(11\bar{2}0)$-plane/$(0001)$-plane hematite single crystals. The scale bar is 100 μm. (b)/(d) Two repeatable normalized polar/longitudinal MOKE intensities are achieved by applying a square wave of $H_y$/$H_x$ with an amplitude of 1 Oe/2 Oe at a frequency of 0.1 Hz. The white circles in the corresponding MOKE images mark the regions of interest. The scale bar in (b)/(d) is 10 μm/25 μm, respectively.

Similarly, altermagnetic domains in the $(0001)$-plane hematite can be studied using the longitudinal MOKE imaging experiment. In the presence of $H_x$, a clear longitudinal MOKE contrast is obtained in Fig. 5(c), which reflects the difference between two altermagnetic domains of opposite Néel vectors $L_{+y}$ and $L_{-y}$. Under a square-wave driving field $H_x = \pm 2$ Oe with a frequency of 0.1 Hz, the periodic change of time-dependent longitudinal MOKE signals also witnesses the reversible and deterministic switching between $L_{+y}$ and $L_{-y}$, as shown in Fig. 5(d). Thus, our experiments suggest that standard MOKE imaging microscopy can be conveniently employed to investigate domain and DW dynamics in altermagnets.

**Conclusions**

We have both theoretically and experimentally studied the symmetry-driven magneto-optical Kerr effect (MOKE) in the altermagnet candidate $\alpha$-$Fe_2O_3$ at room temperature. Through setting orientations of Néel vectors ($L$) by external magnetic fields, we have studied the connection between magnetic space groups (MSG) and the corresponding MOKE responses. In particular, we have found that the $(11\bar{2}0)$-plane hematite exhibits a pronounced spontaneous polar Kerr rotation angle under Néel vector $L \parallel x$, which is enabled by the MSG $C2'/c'$. By contrast, the $(0001)$-plane hematite exhibits a spontaneous longitudinal Kerr rotation angle and the absence of the polar MOKE under Néel vector $L \parallel y$, which is enabled by the MSG $C2/c$. The connection between the MOKE responses and orientations of Néel vectors suggests the importance of magnetic crystal symmetry and altermagnetic nature of hematite. By using the giant Kerr rotation angle in hematite, we further demonstrated that the standard MOKE imaging microscopy can be conveniently used to study the field-driven dynamics of altermagnetic domains and domain walls. These findings not only reinvigorate the foundation of altermagnetism but also suggest the feasibility of studying domain wall physics and devices in altermagnets by optical methods, which could excite more opportunities for the magnetism community.

Note added: While submitting our manuscript, we are aware of a work by H. Pan *et al* [73] has similar conclusions.

# Methods

## Sample information and magnetization characterization

The $(11\bar{2}0)$- and $(0001)$-plane hematite single crystals with the dimensions of 5 mm ×5 mm × 0.5 mm are commercially available from SurfaceNet GmbH (Germany). Magnetization characterization was conducted by the vibrating sample magnetometer.

## Magneto-optical Kerr rotation angle measurements

Polar MOKE measurements were carried out with the linearly polarized light incident normal to the $(11\bar{2}0)$- and $(0001)$-plane samples, by using a home built MOKE magnetometer. The resulting polar Kerr rotation angle of the reflected light were recorded as a function of the applied magnetic field. The measurements are performed with a spot size ∼ 80 μm, and the photon energy ranging from 2.60 eV to 3.00 eV in the wavelength-dependent measurements in Fig. 3(d). Longitudinal MOKE measurements were performed with linearly polarized light at an oblique incidence on the (0001)-plane sample surface, and the resulting Kerr rotation of the reflected light as a function of the applied magnetic field was recorded.

## Imaging and field-driven altermagnetic domains

Polar and longitudinal MOKE images of domains in the $(11\bar{2}0)$- and (0001)-plane hematite are captured by using a commercial MOKE microscope (evico magnetics GmbH, Germany). Deterministic and reversible switching between altermagnetic domains with opposite Néel vectors was achieved by applying a periodic square-wave magnetic field with an amplitude of 1 Oe/2 Oe at 0.1 Hz in the $(11\bar{2}0)$-/(0001)- plane hematite.

## The First-principles calculations

The electronic structure calculations were carried out using the full-potential linearized augmented plane-wave (FP-LAPW) method as implemented in the FLEUR code [74]. The GGA-PBE exchange-correlation functional and spin-orbit coupling (SOC) were considered in all calculations. An energy cutoff of 4.6 $a_0^{-1}$ (where $a_0^{-1}$ is the Bohr radius) and an energy convergence criterion of $10^{-5}$ Ha were adopted. The Brillouin zone was sampled using a 12×12×12 k-point mesh. To properly account for the on-site Coulomb interaction of Fe 3*d*-orbitals, the GGA+U scheme was applied with a Hubbard parameter $U = 5.5$ and Hund's coupling $J = 0.5$ eV [75]. Experimental lattice constants ($a$ = 5.035 Å, $c$=13.758 Å) were used throughout the calculations [53]. The maximally-localized Wannier functions (MLWFs) for Fe *d* and O *p* orbitals were constructed using the WANNIER90 package [76], based on the converged electronic structure. These MLWFs were then employed to evaluate the optical conductivity tensor on an ultra-dense *k*-point grid of 201×201×201 [77]. To examine the influence of weak net magnetization on the Kerr effect, the Néel vector was constrained along the [1$\bar{1}$00] direction, and with and without a small canting angle of 0.065° toward the [11$\bar{2}$0] direction, respectively.

## Data availability

All data that support the findings of this study are available from the corresponding authors on reasonable request.

## Acknowledgments

Work carried out at Tsinghua is supported by the distinguished Young Scholar program of National Natural Science Foundation of China (NSFC Grant No. 12225409), Beijing Natural Science Foundation (Grant No. Z240006), NSFC Basic Science Center Project (NSFC Grant No. 52388201), the National Key R&D Program of China (Grant No. 2022YFA1405100), the NSFC general program (NSFC Grant Nos. 52271181, 12421004, 12404138), and the Innovation Program for Quantum Science and Technology (Grant No. 2023ZD0300500). L. Yang and J. Liang are supported by the NSFC (NSFC Grant No. 12361141826). Y. Yao is supported by the NSFC (NSFC Grant Nos. 12321004, W2511003).

## Author contributions

J. Luo, J. Liang, L. Y. and W. J. initiated and designed the experiment. J. Luo conducted the MOKE images and magnetization measurements. X. Z. and Y. Y. conducted the first-principles calculations. J. Liang conducted the Kerr rotation angles measurements. L. W., Q. Z., Y. J. and W.W. contributed to experimental discussions and manuscript revision. J. Luo. and W. J. wrote the manuscript. L. Y. and W. J. supervised the project. All authors discussed the results and approved the final manuscript.

## Competing interests

The authors declare no competing interests.